\documentstyle[preprint,aps]{revtex}
\newcommand{\doublespace}{\renewcommand{\baselinestretch}{1.5}\large\normalsize}
\newcommand{\BE}{\begin{equation}}
\newcommand{\EE}{\end{equation}}
\newcommand{\BA}{\begin{eqnarray}}
\newcommand{\EA}{\end{eqnarray}}
\begin{document}
\draft 
\tightenlines
\doublespace

\title{Noise-induced flow in quasigeostrophic 
turbulence \\
with bottom friction}

\author{Alberto Alvarez, Emilio Hern\'andez-Garc\'\i a, 
 and Joaqu\'\i n Tintor\'e}
\address{
Departament de F\'\i sica \\
and\\
Instituto Mediterr\'aneo de Estudios Avanzados (IMEDEA\cite{imedea}, 
CSIC-UIB)\\  
Universitat de les Illes Balears\\
E-07071 Palma de Mallorca, Spain}
\date{\today}
\maketitle

\begin{abstract}
Randomly-forced fluid flow in the presence of scale-unselective 
dissipation  develops mean currents following topographic 
contours. Known mechanisms based on the scale-selective action of 
damping processes are not at work in this situation. 
Coarse-graining reveals that the phenomenon is a kind of 
noise-rectification mechanism, in which lack of detailed balance
and the symmetry-breaking provided by topography  
play an important role. 

\end{abstract}


\newpage
\doublespace
 In the last decades much effort has been devoted towards 
understanding fluid motion in situations in which vertical 
velocities are small and slaved to the horizontal motion. Under 
these circumstances the flow can be described in terms of two 
horizontal coordinates, the vertical depth of the fluid becoming a 
dependent variable. The fluid displays many of the unique 
properties of two-dimensional turbulence, but some of the aspects 
of three-dimensional dynamics are still essential. Among the 
several quasi two-dimensional dynamics considered, barotropic 
quasigeostrophic turbulence \cite{williams78,pedlosky87} has 
focussed most of the interest.  The reason for such interest lies 
on the relevance of this dynamics as a model to understand 
planetary atmospheres and ocean general circulations as well as to 
describe some plasma physics phenomena\cite{mima78,hasegawa79,hasegawa85}. 
In 
the first case the twodimensionality of the flow is induced by the 
Coriolis force, whereas in the second it may arise from the action 
of magnetic fields. In the geophysical context, the topography 
over which the layer of fluid flows is the ingredient introducing 
a difference with respect to purely twodimensional turbulence. 

The first works on topographic turbulence focused on its 
statistical properties in the absence of dissipation and forcing 
\cite{salmon76}. These studies established  a tendency of the flow 
to reach a maximum entropy (Gibbs) state characterized by the 
existence of stationary mean currents following isobaths. Mean 
currents associated with topography are however not restricted to 
this inviscid and unforced case, as shown by 
\cite{bretherton76}, where viscosity was included. 
Statistical-mechanics 
equilibrium arguments can not be invoked in this situation of 
decaying turbulence to explain the appearance of  mean currents 
correlated to topographic features. The explanation put forward 
was that viscous damping, due to its stronger action at the 
smallest scales, dissipates enstrophy much faster than energy, a 
quantity more concentrated at large scales in twodimensional 
turbulence settings. As a consequence of this scale-selective 
behavior, the decaying turbulent state would be the one with the 
smaller enstrophy compatible with the (slowly decreasing in time) 
instantaneous value of the energy. It turns out that these minimum 
enstrophy states are closely related to the equilibrium maximum 
entropy ones and, as them, present distinct mean currents 
following isobaths. The natural tendency of topographic 
turbulence to generate currents along 
isobaths, under different situations,
aroused a significant interest in the geophysical community. 
Specifically, studies were addressed to investigate the role 
played by these currents on the general circulation of the 
world's ocean 
\cite{holloway92,alvarez94,eby94,sou96}. 

It has been recently shown numerically \cite{alvarez97} that mean 
currents following isobaths appear also in situations in which a 
scale-unselective dissipation, Rayleigh friction modeling an 
Eckman bottom drag, is used. This fact was observed in numerical 
simulations in which random forcing (also of a scale unselective 
nature) was present. Thus the state obtained is not one of 
decaying turbulence but a nonequilibrium statistically steady 
state. Statistical properties are not far from those of a 
generalized canonical equilibrium, but this is just a numerical 
coincidence, since neither the maximum entropy nor the minimum 
enstrophy mechanisms, the ones invoked in the previous situations, 
are here at work. Instead \cite{alvarez97} interpreted these 
currents as a noise rectification phenomenon\cite{rectification}: 
the mean currents were generated by nonlinearity and sustained by 
noise, with topography providing the symmetry-breaking ingredient 
giving a particular sense to the flow. Later it was theoretically 
shown and numerically confirmed that noise rectification also 
appears in randomly forced topographic turbulence when a scale 
selective damping (viscosity) is considered 
\cite{alvarez98,alvarez99}. However, the detailed  understanding 
of the noise-rectification 
mechanism when a scale-unselective damping is present
is still an open 
question. Addressing this question
as well as to provide a rigorous theoretical framework to explain 
the numerical results shown in \cite{alvarez97} are the main 
motivation of this Letter. 
 
In the quasigeostrophic approximation, the motion of a single 
layer of fluid in a rotating frame or planet can be described in 
terms of the horizontal components of the velocity $\left( u({\bf 
x}),v({\bf x}) \right)$ that can also be written in terms 
of a streamfunction verifying: 
  
\begin{equation}
u = - {\partial \psi \over \partial y}\ , \ \ \ 
v =  {\partial \psi \over \partial x}
\end{equation}

The streamfunction $\psi({\bf x},t)$, with ${\bf x} \equiv (x,y)$, is 
governed by the dynamics \cite{pedlosky87}:  

\begin{equation}
\label{eq}
{\partial \nabla^2 \psi \over \partial t} + 
 \lambda \left[ \psi , \nabla^2 \psi +h \right] =
-\epsilon \nabla^2 \psi + F \ .
\end{equation}

$\epsilon$ is the friction parameter, modelling bottom friction, 
$F({\bf x},t)$ is any kind of relative-vorticity external forcing, 
and  $h=f \Delta H/H_0$, with $f$ the Coriolis parameter,  $H_0$ 
the mean depth, and $\Delta H({\bf x})$ the local deviation from 
the mean depth. $\lambda$ is a bookkeeping parameter introduced to 
allow perturbative expansions in the interaction term. The 
physical case corresponds to $\lambda=1$.  The Poisson bracket or 
Jacobian is defined as 
\begin{equation}
\label{jac}
[A,B]= {\partial A \over \partial x} {\partial B \over \partial y} - 
{\partial B \over \partial x}{\partial A \over \partial y}\ . 
\end{equation}

Equation (\ref{eq}) gives the time evolution of relative vorticity 
subjected to forcing and dissipation. The form used by the 
dissipation term acts with the same strength at all spatial 
scales, in contrast with viscosity terms of the form $\nabla^4 
\psi$, excluded from our model (\ref{eq}), which would dissipate 
better the smaller scales. A convenient choice of $F$, able to 
model a variety of processes, is to assume it to be a Gaussian 
stochastic process with zero mean, and with a Fourier transform 
$\hat F_{\bf k}(\omega)$ having the following two-point 
correlation function: 
\begin{equation}
\left< \hat F_{\bf k}(\omega)\hat F_{\bf k'}(\omega' ) \right> = D 
k^{-y} \delta ({\bf k}+{\bf k'})\delta (\omega+\omega') \ .
\end{equation}  
Here, ${\bf k}=(k_x,k_y)$, and $k=|{\bf k}|$. Thus 
the process is white in time but has power-law correlations in 
space.  Stochastic forcing has been used in fluid dynamics 
problems to model stirring forces\cite{marti97}, short scale 
instabilities\cite{williams78}, thermal noise \cite{treiber96,ll}, 
or processes below the resolution of computer models 
\cite{mason94}, among others \cite{mccomb95}. 

Guided by the results in \cite{alvarez98}, we attribute the 
average currents to a large-scale rectification of the small-scale 
fluctuations introduced by the noise term. In consequence we 
resort to a coarse-graining procedure to investigate how the 
dynamics of long-wavelength modes in (\ref{eq}) is affected by the 
small scales.  For our problem it is convenient to use the Fourier 
components of the streamfunction $\hat \psi_{{\bf k}\omega}$ or 
equivalently the relative vorticity $\zeta_{{\bf k}\omega}=-k^2 
\hat\psi_{{\bf k}\omega}$. This variable satisfies: 
\begin{eqnarray}
\label{fourier}
\zeta_{{\bf k}\omega} &=& G^{0}_{{\bf k}\omega} F_{{\bf k}\omega} +  
                                            \nonumber \\
&&\lambda G^{0}_{{\bf k}\omega} 
\sum_{{\bf p},{\bf q},\Omega,\Omega'} A_{{\bf k}{\bf p}{\bf q}} 
\left(   \zeta_{{\bf p}\Omega} \zeta_{{\bf q}\Omega'} + 
        \zeta_{{\bf p}\Omega} h_{\bf q}                  \right)\ \ ,
\end{eqnarray}
where the interaction coefficient is:
\begin{equation}
\label{interaction}
A_{{\bf k}{\bf p}{\bf q}}= 
(p_x q_y - p_y q_x) p^{-2} \delta_{{\bf k},{\bf p}+{\bf q}}\ \ , 
\end{equation}
the bare propagator is:
\begin{equation}
\label{propagator} 
G^{0}_{{\bf k}\omega}=(-i\omega + \epsilon)^{-1}\ \ , 
\end{equation}
and the sum is restricted by ${\bf k}={\bf p}+{\bf q}$ and 
$\omega=\Omega+\Omega'$. ${\bf p}=(p_x,p_y)$, $p=|{\bf p}|$, and 
similar  expressions hold for ${\bf q}$. The wavenumbers are 
restricted to $0<k<k_0$, with $k_0$ an upper cut-off. Following 
the method in Ref. \cite{turbulence2d}, one can eliminate the 
modes $\zeta^{>}_{{\bf k}\omega}$ with $k$ in the shell $k_0 
e^{-\delta}< k < k_0 $ and substitute their expressions into the 
equations for the remaining low-wavenumber modes $\zeta^{<}_{{\bf 
k}\omega}$ with $0 < k < k_0 e^{-\delta}$. The parameter $\delta$ 
measures thus the width of the band of eliminated wavenumbers. To 
second order in $\lambda$, the resulting equation of motion for 
the modes $\zeta^{<}_{{\bf k}\omega}$, written in terms of the 
large-scale streamfunction $\psi^<({\bf x},t)$ is: 
\begin{eqnarray}
\label{result}
{\partial \nabla^2 \psi^{<} \over \partial t} &+& \lambda \left[ 
\psi^{<} , \nabla^2 \psi^{<} +h^{<} \right] =   \nonumber 
\\ 
&-&\epsilon \nabla^2 \psi^{<} - g \nabla^4 h^{<} + \nu'  \nabla^4 
\psi^{<}+ F' \ , 
\end{eqnarray}
where 
\begin{equation}
\label{nu}
\nu'=  {\lambda^2 S_2 D  y  \delta \over 16 (2 \pi)^2 \epsilon^2} 
, 
\end{equation}
\begin{equation}
\label{g}
g(\lambda, D, \delta, \epsilon, y)= 
{\lambda^2 D S_2 (y+2) \delta \over 32 (2 \pi)^2 \epsilon^2}\ \ .
\end{equation}

$F'({\bf x},t)$ is an effective noise which turns out to be also a 
Gaussian process but with mean value and correlations given by: 
\begin{equation}
\label{average}
<F'({\bf x},t)>=-{ \lambda^2 D S_2 (2+ y )\delta  \over 32 (2 \pi)^2 \epsilon^2} 
\nabla^4 h^{<},
\end{equation}
\begin{eqnarray}
\label{correlations}
\left< 
      \left(  \hat F'_k(\omega)-\left< \hat F'_k(\omega) \right>
                              \right) 
      \left(   \hat F'_{k'}(\omega')-\left< \hat F'_{k'}(\omega') \right>
                              \right)  
\right> =  \nonumber \\
D k^{-y} \delta (k+k')\delta (\omega+\omega')
\end{eqnarray} 
$S_2$ is the length of the unit circle: $2\pi$. Equations 
(\ref{result})-(\ref{correlations}) are the main result in this 
Letter.  They give the dynamics of long wavelength modes 
$\psi_{{\bf k}\omega}^{<}$ with $0 < k < k_0 e^{-\delta}$. They 
are valid for small $\lambda$ or, when $\lambda \approx 1$, for 
small width $\delta$ of the elimination band. 

The elimination of the small scales leads, as expected from 
physical grounds, to the appearance of an effective viscosity 
$\nu'$ at large scales.  Depending on the sign of $y$, this 
viscous term can be destabilizing. One should keep in mind however 
that Eq.(\ref{result}) is only valid at large scales, so that such 
small-wavelength instabilities, when formally present, would be 
avoided by proper choice of the wavenumber cut-off. More  
importantly, a new term depending on the topography $g\nabla^4  
h^{<}$ has been generated by the small scales. Another similar 
term is contained in the mean value of the effective large-scale 
noise $F'$. Both terms have the effect of pushing the large-scale 
motion towards a state of flow following the isolevels of bottom 
perturbations $h^{<}$.   The energy in this preferred state is 
determined by the function $g(\lambda, D, \delta, \epsilon, y)$ 
which measures the influence of the different terms of the 
dynamics (nonlinearity, noise, friction). From relation (\ref{g}) 
it is clear that while nonlinearities and noise increase the 
intensity of the mean currents in the preferred state, high values 
of the friction parameter would reduce them. 

As in the situation with scale selective damping 
\cite{alvarez98,alvarez99}, (\ref{g}) and (\ref{average}) exhibit 
a characteristic dependence on the exponent $y$ of the 
random-forcing spectral power-law. In the present case, the 
dependence with $y+2$ implies that the directed currents reverse 
sign as $y$ crosses the value $-2$, and that they vanish if 
$y=-2$. It is easy to check that the conditions for detailed 
balance between the fluctuation input and dissipation 
\cite{gardiner} are satisfied precisely for $y=-2$ (in general, 
the condition is $y=-n$, where $n$ is the exponent of $\nabla$ in 
the dissipation term). In this situation the steady-state 
probability distribution for $\psi$ can be found exactly and is 
independent of topography. Thus the generation of flow along 
isobaths is a consequence of lack of detailed balance, a general 
feature of noise-rectifying mechanisms\cite{rectification}.  

Concluding, relations (\ref{result})-(\ref{correlations}) show 
that the origin of average circulation patterns in 
quasigeostrophic turbulence over topography is related to  
nonlinearity and lack of detailed balance. The present approach 
provides a theoretical basis for the numerical observations 
presented in \cite{alvarez97}, for which mechanisms based on the 
scale-selective action of damping can not be applied. Nonlinear 
terms couple the dynamics of small scales to the large ones and 
provide the  mechanism for energy transfer from the fluctuating 
component of the spectrum to the mean one. This mean spectral 
component, absent in purely two-dimensional turbulence 
\cite{thom}, is controlled by the shape of the bottom boundary, 
which breaks the isotropy of the system, and characterizes the 
structure of the flow pattern.

Financial support from CICYT (AMB95-0901-C02-01-CP and 
MAR98-0840), DGICYT (PB94-1167), and from the MAST program MATTER 
MAS3-CT96-0051 (EU) is greatly acknowledged.


\newpage


\begin{references}

\bibitem[\dagger]{imedea} URL: {\tt http://www.imedea.uib.es/}

\bibitem{williams78} G. P. Williams,
J. Atmos. Sci. 35 (1978) 1399.

\bibitem{pedlosky87} J. Pedlosky, Geophysical fluid dynamics,
(Springer-Verlag, New York, 1987).

\bibitem{mima78} A. Hasegawa, K. Mima, Phys. Fluids
21 (1978) 87.

\bibitem{hasegawa79} A. Hasegawa, C. G. Maclennan,
Phys. Fluids 22 (1979) 11.

\bibitem{hasegawa85} A. Hasegawa, 
Advances in Physics 34 (1985) 1.

\bibitem{salmon76} R. Salmon, C. Holloway, M.G. Hendershot, J. 
Fluid Mech 75 (1976) 691. 

\bibitem{bretherton76} F.P. Bretherton, D.B. Haidvogel, J. Fluid 
Mech. 78 (1976) 129. 

\bibitem{holloway92} G. Holloway,
J. Phys. Oceanogr. 22 (1992) 1033.

\bibitem{alvarez94} A. \'Alvarez, J. Tintor\'e, 
G. Holloway, M. Eby, J.M. Beckers,
J. Geophys. Res. 99 (1994) 16053.

\bibitem{eby94} M. Eby, G. Holloway,
J. Phys. Oceanogr. 24 (1994) 2577.

\bibitem{sou96} T. Sou, G. Holloway, M. Eby
J. Geophys. Res. 101 (1996) 16449.

\bibitem{alvarez97} A. \'Alvarez, E. Hern\'andez-Garc\'\i a, J.
Tintor\'e,
Physica A 247 (1997) 312. {\tt chao-dyn/9701009}

\bibitem{rectification}
F. J\"{u}licher, A. Ajdari, and J. Prost, Rev. Mod. Phys. 69, 1269 
(1997); M.O. Magnasco, Phys. Rev. Lett. 71 (1993) 1477; J. 
Maddox, Nature 369 (1994) 181; C. R. Doering, W. 
Horsthemke, J. Riodan, J. Phys. Rev. Lett. 72 (1994) 2984; 
J. Maddox, Nature 368 (1994) 287; J.K. Douglass, Lon 
Wilkens, Eleni Pantazelou, Frank Moss, Nature 365 (1993) 337; 
S. M. Bezrukov, Igor Vodyanoy, Nature 378 (1995) 362; 
J. Rousselet, L. Salome, A. Ajdari, J. Prost,  Nature 
370 (1994) 446. 


\bibitem{alvarez98} A. \'Alvarez, E. Hern\'andez-Garc\'\i a, J. Tintor\'e,
Phys. Rev. E 58 (1998) 7279. {\tt chao-dyn/9802003}

\bibitem{alvarez99} A. \'Alvarez, E. Hern\'andez-Garc\'\i a, J. Tintor\'e,
in preparation. 

\bibitem{marti97} A.C. Mart\'\i , J.M. Sancho, F. Sagu\'es, and A. Careta, 
Phys. Fluids 9 (1997) 1078. {\tt chao-dyn/9703015}
  
\bibitem{treiber96} M. Treiber, Phys. Rev. E 53 (1996) 577. 

\bibitem{ll} L.D. Landau and M. Lifshitz, Fluid Dynamics, 2nd Edition.
Course of Theoretical Physics vol. 6 (Pergamon, New York, 1987). 

\bibitem{mason94} P.J. Mason, 
Q.J.R. Meteorol. Soc. 120 (1994) 1. 

\bibitem{mccomb95} W.D. McComb, 
Rep. Prog. Phys. 58 (1995) 1117.

\bibitem{turbulence2d} D. Foster, D. Nelson, M. Stephen,
Phys. Rev. A 16 (1977) 732. 

\bibitem{gardiner} C. W. Gardiner, Handbook of Stochastic
methods for Physics, Chemistry and the Natural Sciences.
Springer, New York, 1989.

\bibitem{thom} P.D. Thompson,
J. Fluid Mech. 55 (1972) 711. 
 

\end{references}
\end{document}